\documentclass[preprint,amsmath,amssymb,nofootinbib,nobibnotes,superscriptaddress, prb, 11 pt]{revtex4-1}

\usepackage{graphicx}
\usepackage{dcolumn}
\usepackage{amsmath}
\usepackage{bm}
\usepackage[utf8]{inputenc}
\usepackage{mathtools}
\usepackage{pgffor}
\usepackage[normalem]{ulem}
\usepackage{xcolor}
\usepackage{pdfpages}
\usepackage{BOONDOX-cal}
\makeatletter

\AtBeginDocument{\let\LS@rot\@undefined}
\makeatother
\def\supplementfilename{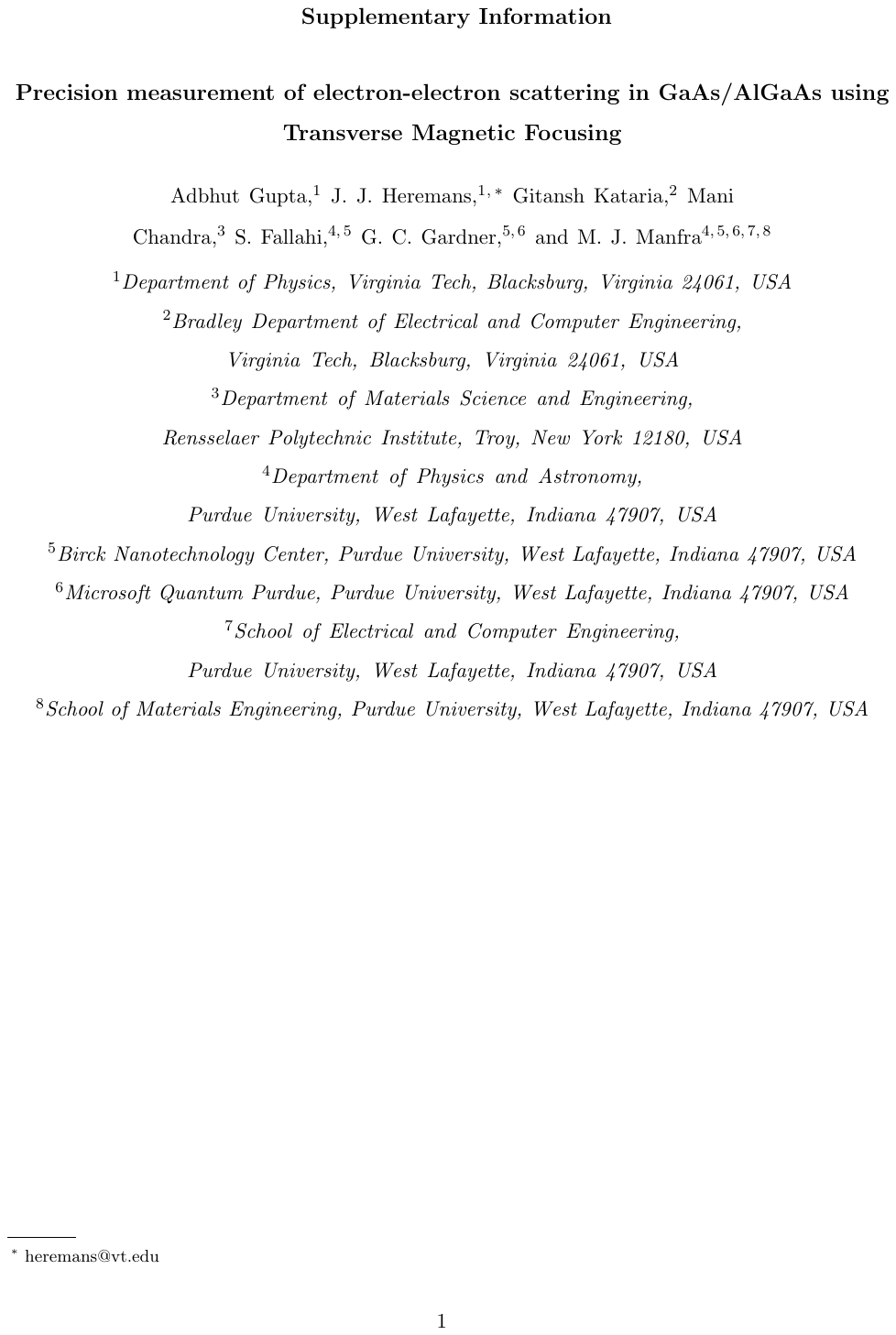}
\pdfximage{\supplementfilename}
\def\numbersupplementpages{\the\pdflastximagepages}
\newif\ifarXiv
\arXivtrue 

\begin{document}

\title{Precision measurement of electron-electron scattering in GaAs/AlGaAs using transverse magnetic focusing}

\author{Adbhut Gupta}
\affiliation{Department of Physics, Virginia Tech, Blacksburg, Virginia 24061, USA}
\author{J. J. Heremans}
\email{heremans@vt.edu}
\affiliation{Department of Physics, Virginia Tech, Blacksburg, Virginia 24061, USA}
\author{Gitansh Kataria}
\affiliation{Bradley Department of Electrical and Computer Engineering, Virginia Tech, Blacksburg, Virginia 24061, USA}
\author{Mani Chandra}
\affiliation{Department of Materials Science and Engineering, Rensselaer Polytechnic Institute, Troy, New York 12180, USA}
\author{S. Fallahi}
\affiliation{Department of Physics and Astronomy, Purdue University, West Lafayette, Indiana 47907, USA}
 \affiliation{Birck Nanotechnology Center, Purdue University, West Lafayette, Indiana 47907, USA} 
\author{G. C. Gardner}
 \affiliation{Birck Nanotechnology Center, Purdue University, West Lafayette, Indiana 47907, USA} 
 \affiliation{Microsoft Quantum Purdue, Purdue University, West Lafayette, Indiana 47907, USA}
\author{M. J. Manfra}
\affiliation{Department of Physics and Astronomy, Purdue University, West Lafayette, Indiana 47907, USA}
\affiliation{Birck Nanotechnology Center, Purdue University, West Lafayette, Indiana 47907, USA} 
\affiliation{Microsoft Quantum Purdue, Purdue University, West Lafayette, Indiana 47907, USA}
\affiliation{School of Electrical and Computer Engineering, Purdue University, West Lafayette, Indiana 47907, USA} 
\affiliation{School of Materials Engineering, Purdue University, West Lafayette, Indiana 47907, USA}

\begin{abstract}

Electron-electron (e-e) interactions assume a cardinal role in solid-state physics. Quantifying the e-e scattering length is hence critical. In this paper we show that the mesoscopic phenomenon of transverse magnetic focusing (TMF) in two-dimensional electron systems forms a precise and sensitive technique to measure this length scale. Conversely we quantitatively demonstrate that e-e scattering is the predominant effect limiting TMF amplitudes in high-mobility materials. Using high-resolution kinetic simulations, we show that the TMF amplitude at a maximum decays exponentially as a function of the e-e scattering length, which leads to a ready approach to extract this length from the measured TMF amplitudes. The approach is applied to measure the temperature-dependent e-e scattering length in high-mobility GaAs/AlGaAs heterostructures. The simulations further reveal current vortices that accompany the cyclotron orbits - a collective phenomenon counterintuitive to the ballistic transport underlying a TMF setting. 

\end{abstract}

\maketitle 

Electron-electron (e-e) interactions or scattering play an important role in electronic transport and in solid-state physics in general as they determine the quasiparticle lifetime in a Fermi liquid.  Since e-e scattering conserves the total momentum internal to the system we will refer to the e-e scattering as momentum-conserving (MC) scattering in this work. While not affecting mobility due to conservation of total system momentum, in device geometries constructed using two-dimensional electron systems (2DESs) strong MC scattering leads to hydrodynamic phenomena such as vortices\cite{Gurzhi, deJong, govorov, levitov, kumarsuperballistic, shytov, bandurin2018, Lucas, chandra2019, chandraquantum, Gupta2021, Pellegrino2016, Torre2015, Polini2020}. The MC scattering time-scale $\tau_{\textrm{MC}}$, as a fundamental quantity in Fermi liquid theory, has been the subject of several calculations \cite{GQ1982, Zheng1996, Qian2005, Li2013}. However, direct measurements of $\tau_{\textrm{MC}}$ have been more elusive, only so far achieved at lower temperature $T$ ($\lesssim 4 \, \text{K}$) by loss of quantum interference \cite{Yacoby, Lin2002, HeremansPRB98}, tunneling measurements \cite{Murphy}, or scattering measurements \cite{MolenkampSST1992, Jura}. At higher $T \, \gtrsim 4 \, \text{K}$, the effect of MC scattering is to impose a local thermal equilibrium, and here experimental measurements of $\tau_{\textrm{MC}}$ have only recently been enabled by the hydrodynamic transport regime \cite{Gupta2021,Keser2021}.  

We present transverse magnetic focusing (TMF) as a sensitive technique for the measurement of the MC scattering length ($\mathcal{l_\textrm{MC}}$) in a 2DES, which for a circular Fermi surface is equivalent to $\tau_{\textrm{MC}}$. The 2DES is hosted by an ultraclean GaAs/AlGaAs heterostructure grown by optimized MBE process, and is well-suited for this study because the very long momentum-relaxing (MR) scattering length ($\mathcal{l_\textrm{MR}}=$ mobility mean-free path $\simeq 65$\ $\mu$m at $T$ = 4.2 K) provides high sensitivity to MC scattering \cite{Gupta2021}. MR scattering (mediated by lattice defects and phonons) is responsible for dissipation of the system's momentum to the lattice. TMF is usually applied to study the ballistic nature of carriers and characterize the shape of the Fermi surface in 3D solids \cite{Tsoi1974, Tsoi1992} and in 2DESs \cite{Houten1989, Heremans1992, Heremans1999, Taychatanapat, Lee2016, Berdyugin2020}. In the presence of a magnetic field $B$ applied normal to the plane of the 2DES, electrons injected from a point contact (PC) follow semiclassical skipping cyclotron orbits of diameter $d_c = 2\hbar k_F/eB$ to focus on a collector PC at a distance $L_c = n d_c$, where $n$ is an integer. Here $L_c$ represents the center-to-center distance between injector and collector PCs, $k_F$ the Fermi wavevector, $e$ the electron charge, and $\hbar$ the Planck’s constant. The nonlocal resistance, defined as the voltage developed at the collector normalized by the injected current, displays maxima at integer $n$ due to electrons focusing on the collector. The nonlocal resistance at the maxima, denoted by $R_{n}$, is largest when all scattering is absent. The decay in $R_{n}$ with increasing scattering has been observed to be exponential and to possess a characteristic decay length \cite{SpectorSS1990, Hornsey1993}, in early work associated with $\mathcal{l_\textrm{MR}}$ (with no reference to $\mathcal{l_\textrm{MC}}$). The increase in scattering can be effectuated by increasing $T$ as has been noted in graphene \cite{Lee2016, Berdyugin2020}. In this work, we address the problem to relate MC scattering and specifically $\mathcal{l_\textrm{MC}}$ to the decay in $R_{n}$, simultaneously underlining the importance of MC scattering in ballistic transport and introducing a sensitive approach to quantify $\mathcal{l_\textrm{MC}}(T)$. 

Using experimentally-backed extensive high-resolution kinetic simulations wherein $\mathcal{l_\textrm{MC}}$ is an input parameter, we show that for very long $\mathcal{l_\textrm{MR}}$ the decay in the first peak amplitude ($R_{n=1}$) due to MC scattering is universal (independent of geometry/device parameters) and obeys 
\begin{align}
R_{n=1}(d_c, \mathcal{l_\textrm{MC}}) = R_{n=1}(d_c, \infty) - \Delta_{n=1}(d_c) \biggl(1 - \exp\left(-\frac{\alpha d_c}{\mathcal{l_\textrm{MC}}}\right)\biggr)
\label{eq:decay_l_mc}
\end{align}
where $\alpha$ is a dimensionless universal parameter characterizing the decay (found below as $\alpha = 1.34 \pm 0.1$), $R_{n=1}(d_c, \infty)$ is the $B$-dependent peak amplitude in the ballistic limit ($\mathcal{l_\textrm{MC}} \rightarrow \infty$), and $\Delta_{n=1}(d_c)$ is a $B$-dependent prefactor independent of $\mathcal{l_\textrm{MC}}$, setting the decay due to MC scattering. In accordance with phase-space arguments for temperature scaling of e-e interaction in Fermi liquid theory which lead to $\mathcal{l_\textrm{MC}} \propto T^{-2}$, we write: 
\begin{align}
\mathcal{l_\textrm{MC}}(T) &= \alpha d_c \left(\frac{T_{c}}{T}\right)^{2}
\label{eq:l_mc_T}
\end{align}
where $T_{c}$ denotes the (unknown) temperature at which $\mathcal{l_\textrm{MC}} = \alpha d_c$. Substituting Eq.~(\ref{eq:l_mc_T}) into Eq.~(\ref{eq:decay_l_mc}) we obtain $R_{n=1} \propto \exp(-(T/T_{c})^2)$. Therefore, $T_{c}$ represents the characteristic temperature scale for decay of $R_{n=1}$. TMF measurements at variable $T$ in our large-scale GaAs/AlGaAs devices will confirm the predicted decay in $R_{n=1}$, from which we extract $T_c$, ultimately obtaining $\mathcal{l_\textrm{MC}}(T)$.  We perform measurements in three devices, each containing several collector PCs placed at distinct $L_c$. 

In addition to the known skipping cyclotron orbits along the device boundary, our simulations also reveal accompanying current vortices in the regimes from low to dominant MC scattering. A collective phenomenon such as vortices being observed in a TMF setting, which is widely regarded as a purely ballistic experiment, is counterintuitive. Yet, at $B=0$ it has been shown both theoretically and experimentally that even ballistic dynamics can lead to collective phenomena \cite{chandra2019, chandraquantum, Gupta2021}. Here we show additional evidence in the presence of $B$. 

\begin{figure*}[!t]
\begin{center}
\includegraphics[width=6.6 in]{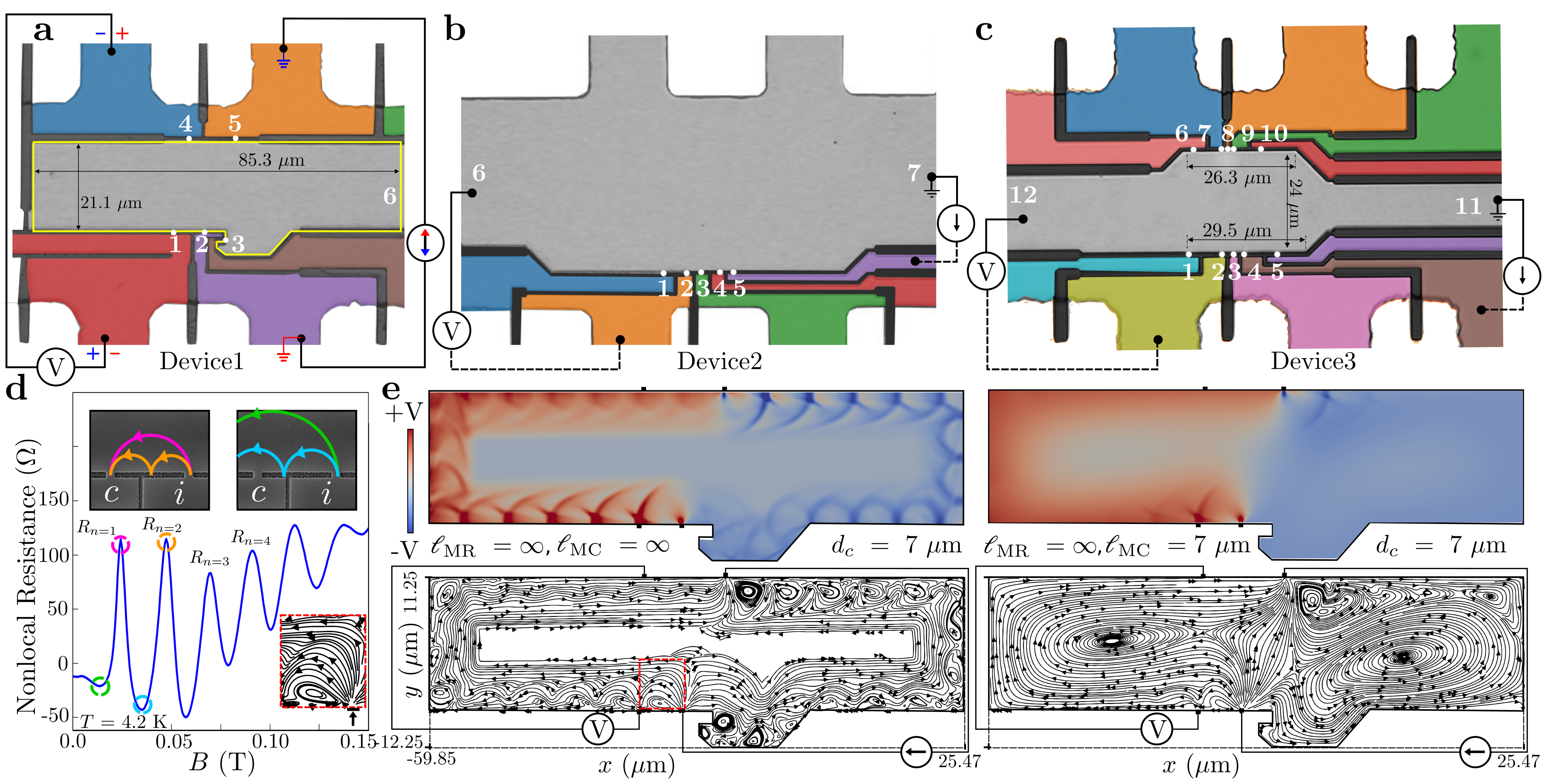}
\caption{\label{fig:fig1}\textbf{Device geometry and TMF spectra. a,b,c,} Optical images of Device1, Device2 and Device3 respectively, showing dimensions, PCs (indicated by white dots), and measurement configurations with current and voltage PCs marked. The paths leading to the PCs are depicted in different colors for distinct visualization.  The computational domain (yellow outline) is indicated for Device1. Device1 features 2 in-line TMF geometries with $L_c=$ 7\ $\mu$m and 10.5\ $\mu$m. Device2 features TMF geometries with 7 distinct $L_c$ ranging from 3\ $\mu$m to 15\ $\mu$m. Device3 features TMF geometries with 10 distinct $L_c$ ranging from 1.3\ $\mu$m to 20.5\ $\mu$m (data from 6 geometries each in Device2 and Device3 are presented, omitting closely spaced $L_c$).  \textbf{d,} TMF spectrum (nonlocal resistance vs $B$ in Device1, $L_c=$ 7\ $\mu$m) $= V_{1,4}/I_{2,5}$ (where $V_{1,4} = V_1 - V_4$  and $I_{2,5}$ is the conventional current from 2 to 5 inside the device) as obtained in experiment at $T$ = 4.2 K. The inset shows schematics of semiclassical cyclotron orbits corresponding to first two maxima and minima, indicated using the same color in the TMF spectrum. Maxima ($R_n$) occur at $L_c = n d_c$ where integer $n$ represents the number of orbits electrons follow before focusing into $c$ ($n$ indicated on the corresponding maxima). Minima occur at values slightly less than half-integer $n$. \textbf{e,} Simulated voltage contour plots (top) and current streamlines (bottom) in Device1 for the first maximum ($d_c =$ 7 $\mu$m) for $\mathcal{l_\textrm{MC}} \rightarrow \infty$ (left) and finite  $\mathcal{l_\textrm{MC}}$ (right). Both show cyclotron orbits and vortices with distinct voltage and current patterns, which for $\mathcal{l_\textrm{MC}} \rightarrow \infty$ are highlighted in the dashed red box and are magnified in the inset of \textbf{(d)} (cfr text: $\mathcal{l_\textrm{MR}} \rightarrow \infty$ throughout most simulations).} 
\end{center}
\end{figure*}

\section*{Results}
\subsection*{TMF devices}

The TMF geometries were patterned on a 2DES in a quantum well in an ultraclean GaAs/AlGaAs heterostructure of mobility $\mu$ exceeding 670 m$^2$V$^{-1}$s$^{-1}$ at $T$ = 4.2 K across all devices (Fig.~\ref{fig:fig1}a). At the areal electron density $N_s \approx 3\times10^{15}$ m$^{-2}$, the Fermi energy $E_F \approx$ 10.9 meV. The methods and transport properties are described in Supplementary Note 1, Note 2. Experiments were performed on multiprobe Hall mesas in three devices $-$ Device1, Device2 and Device3 \cite{Gupta2021} bearing numerous in-line TMF geometries with $L_c$ ranging from $7 - 10.5\ \mu$m in Device1, $3 - 15\ \mu$m in Device2 and $1.3 - 20.5\ \mu$m in Device3 (Fig.~\ref{fig:fig1}a-c). Each TMF geometry features two PCs which can act either as an injector  or collector. The conducting width of each PC is $w \approx$ 0.6\ $\mu$m and the Fermi wavelength, $\lambda_F \approx$ 43 nm implying that $w/(\lambda_F /2) \approx$ 28 spin-degenerate transverse modes contribute to transport, yielding a PC resistance $\approx (h/2e^2)/28$ = 461 $\Omega$. The large number of modes indicate that quantized transport through the PC apertures can be neglected. Measurements were performed in the linear response regime at 4.2 K $< T <$ 36 K, using low frequency lock-in techniques without any DC offsets, and under a small excitation current $I \sim$ 100-200 nA to avoid electron heating. The boundaries of the device were defined by wet etching, resulting in specular scattering at the boundaries \cite{Heremans1999, Chen}.

An example of experimental results is depicted in Fig.~\ref{fig:fig1}d for $L_c=$ 7\ $\mu$m in Device1 depicting the untreated nonlocal resistance at the collector vs $B$ (referred to as TMF spectra) at 4.2 K. The maxima originate from cyclotron orbits impinging in the vicinity of, or directly on, the collector ($c$ in inset of Fig.~\ref{fig:fig1}d). For the minima, the orbits straddle the collector (inset of Fig.~\ref{fig:fig1}d). 

\subsection*{TMF simulations}
We simulate magnetotransport in the actual experimental device geometry using BOLT, a high resolution solver for kinetic theories \cite{bolt, chandra2019, Gupta2021}, which solves the Boltzmann transport equation: 
\begin{equation}
\frac{1}{v_F}\frac{\partial f}{\partial t}  + \left(\frac{\mathbf{p}}{mv_F}\right).\frac{\partial f}{\partial \mathbf{x}} + \left(\frac{2}{d_c}\right)\frac{\partial f}{\partial \theta}  = -\frac{f - {f_0}^{\textrm{MR}}}{\mathcal{l_\textrm{MR}}} - \frac{f - {f_0}^{\textrm{MC}}}{\mathcal{l_\textrm{MC}}}
\label{eq:boltzmann}
\end{equation}
where $f(\mathbf{x}, \mathbf{p}, t)$ is the probability distribution of electrons along the spatial coordinates $\mathbf{x} \equiv (x, y)$ and momentum coordinates $\mathbf{p} \equiv \hbar k_F (\cos(\theta), \sin(\theta))$, where $\theta$ denotes the angle on the Fermi surface, $t$ denotes time, and $v_F$ denotes the Fermi velocity. The Lorentz force due $B$ appears in the third term on the left, which for a circular Fermi surface simplifies to the form shown. Long-range electric fields are not explicitly included, but their effects are accounted for at linear order through a renormalized chemical potential \cite{chandra2019}. The injected particles, with an effective mass $m$, are constrained to remain on the Fermi surface, move at $v_F$ and are injected over all angles following a cos($\theta$) distribution which is maximum perpendicular to the boundary into which the contact is placed \cite{chandraquantum}. The effect of the RHS of Eq.~(\ref{eq:boltzmann}) is a thermalization of carriers to local stationary and drifting Fermi-Dirac distributions, $f_0^{\textrm{MR}}$ and $f_0^{\textrm{MC}}$, due to MR and MC scattering respectively. This is implemented using a dual relaxation time approximation with scattering length scales $\mathcal{l_\textrm{MR}}$ and $\mathcal{l_\textrm{MC}}$. Further details regarding the collision operators are given in Ref.~\cite{chandra2019}. We consider perfectly reflecting device boundaries, with carriers injected by imposing a shifted Fermi-Dirac distribution at the locations of the contacts. 

\begin{figure}[!t]
\includegraphics[width=1\linewidth]{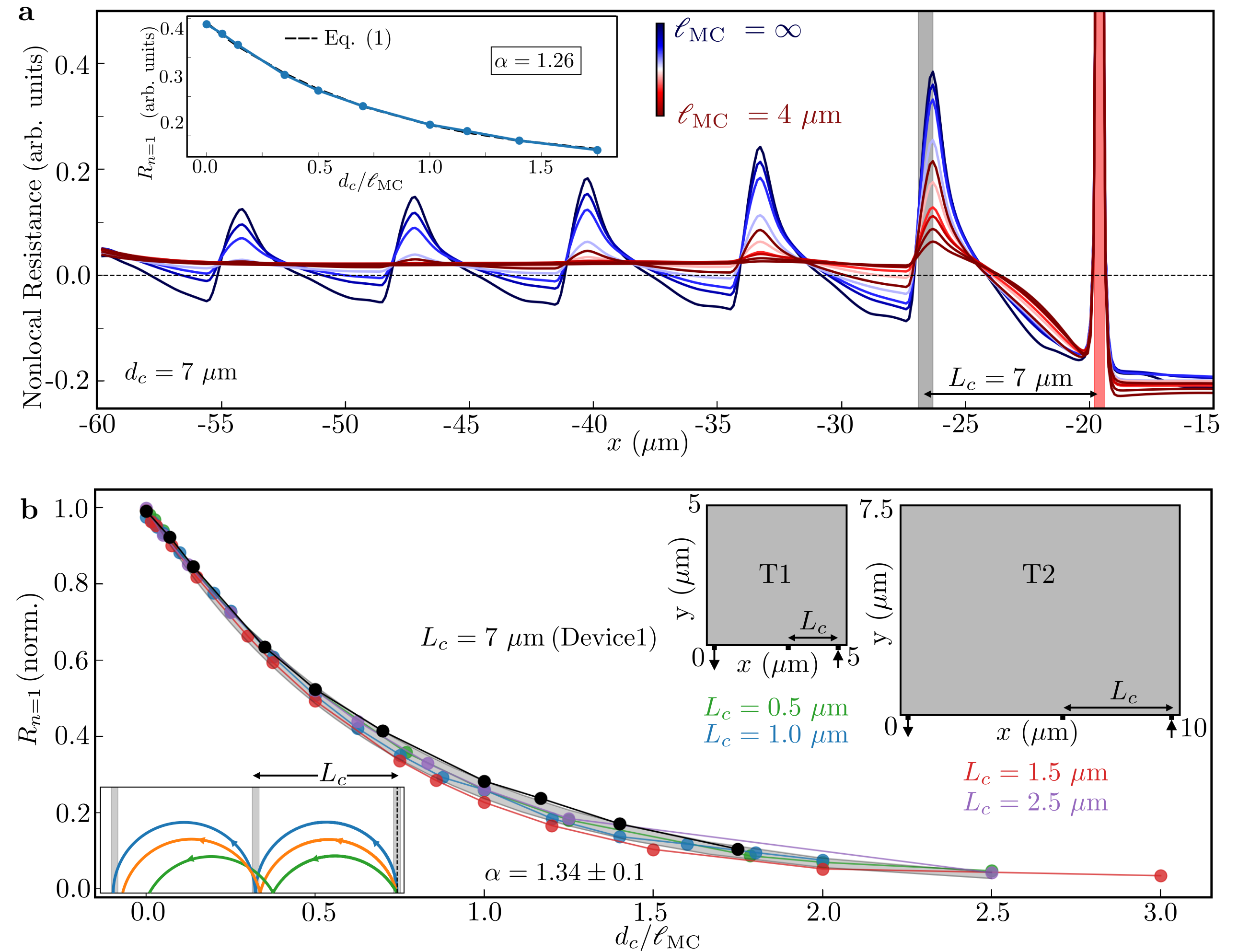}
\caption{\textbf{Universal decay of TMF amplitude. a,} Simulated nonlocal resistance in Device1 (Fig.~\ref{fig:fig1}a,e, $L_c = d_c =$ 7\ $\mu$m) plotted vs position $x$ (corresponding to Fig.~\ref{fig:fig1}e) along the edge of the device into which the injection PC (red vertical bar) is placed, for various values of $\mathcal{l_\textrm{MC}}$ starting from $\mathcal{l_\textrm{MC}} \rightarrow \infty$ to progressively smaller values. The grey bar indicates the position of collector PC. Inset shows the decay of the first maximum as a function of $d_c/\mathcal{l_\textrm{MC}}$, along with a fit to Eq.~(\ref{eq:decay_l_mc}) (black dotted line). \textbf{b,} The normalized $R_{n=1}$ plotted vs parameter $ d_c/\mathcal{l_\textrm{MC}}$ for T1 ($L_c=d_c=0.5,\,1.0\, \mu\text{m}$; variable $\mathcal{l_\textrm{MC}}$), T2 ($L_c=d_c=1.5,\,2.5\, \mu\text{m}$; variable $\mathcal{l_\textrm{MC}}$), and Device1 ($L_c=d_c=$ 7\ $\mu$m; variable $\mathcal{l_\textrm{MC}}$). The normalized curves for all the devices nearly overlap revealing that $R_{n=1}$ follows an exponential decay vs $ d_c/\mathcal{l_\textrm{MC}}$ independent of geometry, with universal $\alpha =$ 1.34 $\pm$ 0.1. Test devices T1 and T2 are depicted as insets with values of $L_c=d_c$ simulated for each device. The evolution of the current streamlines as we vary $\mathcal{l_\textrm{MC}}$ is depicted in Supplementary Note 3 for T1 for $L_c=d_c=1.0\, \mu\text{m}$. The inset depicts the defocusing of electrons injected at angles different from 90$^{\circ}$ (green and orange trajectories) resulting in lesser number of electrons focusing exactly at $n d_c$ with each reflection and in a decrease in maximum amplitude despite specular boundary reflection.}
\label{fig:fig2}
\end{figure}

The simulations are performed in two simplified test geometries T1, and T2, and in the complicated experimental geometries. To initially isolate the effects of $\mathcal{l_\textrm{MC}}$ on TMF spectra, we set $\mathcal{l_\textrm{MR}} \rightarrow \infty$, neglect Fermi surface thermal smearing and treat $\mathcal{l_\textrm{MC}}$ as a free parameter. We show later that the exceptionally long $\mathcal{l_\textrm{MR}}$ in our 2DES has a minimal effect on the TMF spectra and that the effect of Fermi surface smearing on $R_{n=1}$ is negligible. We start in Fig.~\ref{fig:fig1}e, left panels with the limiting ballistic case where $\mathcal{l_\textrm{MC}} \rightarrow\infty$, and consider Device1 with $d_c$ set to $L_c=$ 7 $\mu$m (fixed $B$). As expected from single-particle insight, carriers emanating from the injector propagate along skipping orbits on the bottom edge, under the influence of Lorentz force. Interestingly, current vortices accompany these cyclotron orbits, even in the absence of all microscopic interactions. This observation reinforces the existence of collective phenomena in the ballistic transport regime, as highlighted in recent work \cite{chandra2019, chandraquantum, Gupta2021}. Next, in Fig.~\ref{fig:fig1}e, right panels we approach the hydrodynamic regime by setting $\mathcal{l_\textrm{MC}} = d_c = 7$ $\mu$m, and observe a profound change in the voltage contour and current streamline plots. A large vortex inhabits the main chamber and displaces most of the cyclotron orbits except orbits near the injector. Notably, the vortices in the hydrodynamic regime are distinct from those in the ballistic regime\cite{Gupta2021}. The ballistic regime exhibits multiple vortices of various scales and at various locations in the device even in the absence of electron-electron interactions (see left panel of Fig.~\ref{fig:fig1}e and Supplementary Fig.~4), while the dominance of electron-electron scattering in the hydrodynamic regime favors large device-scale vortices (see right panel of Fig.~\ref{fig:fig1}e and Supplementary Fig.~3). When plotting the simulated nonlocal resistance vs position $x$ along the bottom edge of the device, for fixed $B$ (corresponding to $d_c=L_c=$ 7 $\mu$m) for various $\mathcal{l_\textrm{MC}}$ (Fig~\ref{fig:fig2}a), the role of MC scattering in limiting the TMF signal becomes apparent. Two additional observations appear from Fig~\ref{fig:fig2}a. First, the maxima occur at distances slightly below $L_c = nd_c$, with a deviation of $\lesssim 5$ \% (experimentally leading to TMF maxima slightly below the expected $B= n2\hbar k_F/eL_c$ \cite{Taychatanapat}). Second, $R_n$ decreases with increasing $n$ despite perfect specular boundary scattering. These phenomena result from the angular distribution of injected electrons. The electrons injected at an angle different from 90$^{\circ}$, lead to a defocusing effect such that with each reflection off the boundary, the number of electrons focusing at precisely $nd_c$ decreases, leading to a decrease in $R_n$ with increasing $n$ ($R_{n=1} > R_{n=2} > ...$). This suggests that the ratio of subsequent TMF maxima values $R_n$, is not a good measure to infer specularity of the boundary as has been used by various works \cite{NiheyAPL1990,Lee2016, Berdyugin2020}.

\subsection*{Extraction of $\mathcal{l_\textrm{MC}}$ using TMF}

We next extract the functional dependence of $R_{n=1}$ on $\mathcal{l_\textrm{MC}}$. Varying $\mathcal{l_\textrm{MC}}$ in the simulations, we plot $R_{n=1}$ for the various geometries considered versus the parameter $d_c$/$\mathcal{l_\textrm{MC}}$. In Fig.~\ref{fig:fig2}b we find a data collapse to the function $R_{n=1}(d_c, \mathcal{l_\textrm{MC}}) \propto \exp(-\alpha d_c /\mathcal{l_\textrm{MC}})$, where $\alpha = 1.34 \pm 0.1$ is a dimensionless parameter independent of the device geometry, $B$ or energy dispersion. However, $\alpha$ may depend on the shape of the Fermi surface, restricted in our simulations to a circle. Using the maximal value $R_{n=1}(d_c, \infty)$ in the ballistic limit ($\mathcal{l_\textrm{MC}} \rightarrow \infty$), we obtain the form of Eq.~(\ref{eq:decay_l_mc}), with $\Delta_{n=1}(d_c)$ independent of $\mathcal{l_\textrm{MC}}$. We note that $R_{n=1}$ from Eq.~(\ref{eq:decay_l_mc}) is independent of device geometry because it is a local quantity if $d_c \ll W$ with $W$ the device scale or open distance in any direction away from the PC. For validity of Eq.~(\ref{eq:decay_l_mc}) we require $k_F^{-1} \ll w \ll d_c \ll W$.  The condition $k_F^{-1} \ll w$ ensures that sufficient modes are injected into the system for semiclassical transport. The condition $w \ll d_c$ ensures that magnetic quantization effects are absent. The condition $d_c \ll W$ avoids unwanted boundary scattering, ensuring that skipping orbits reach the collector without hindrance from device boundaries in any direction. We show in the Supplementary Note 4 that when $d_c \sim W$, the decay is no longer universal since carriers now sense the device boundaries. In contrast to $R_n$, the nonlocal resistance at $B=0$ in the ballistic and hydrodynamic regimes as measured in several recent experiments \cite{bandurin2018, Lucas, Gupta2021}, is sensitive to the nonlocal current-voltage relation of the underlying regime, and therefore depends on the device geometry.

\begin{figure}[!t]
\includegraphics[width=1\linewidth]{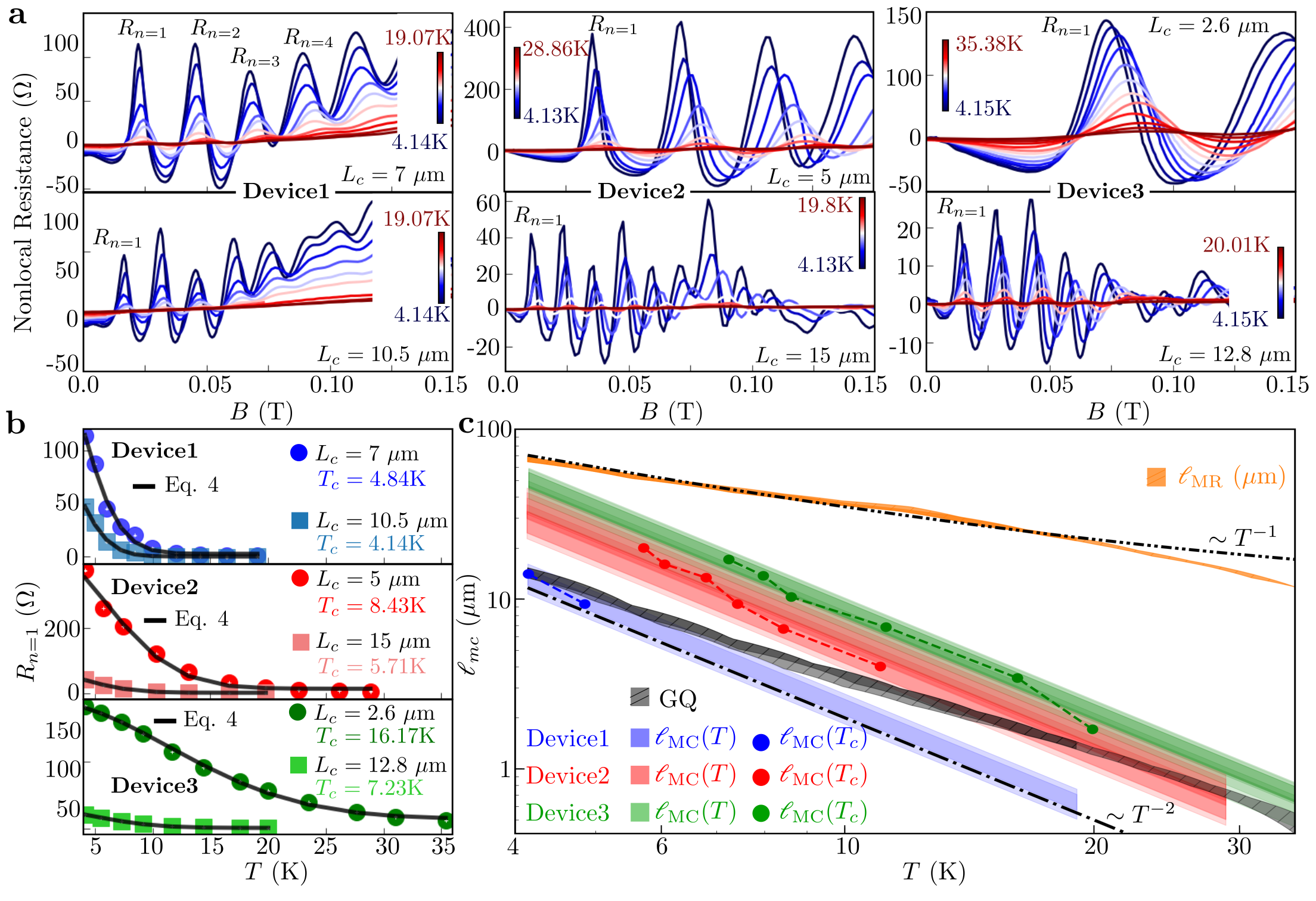}
\caption{\textbf{Dependence on $T$ of TMF and extraction of $\mathcal{l_\textrm{MC}}$. a,} TMF spectra for specified $L_c$ in Device1, Device2 and Device3 over indicated range of $T$ (dependent on $L_c$ and device). Increasing $T$ causes two effects$-$ suppression of TMF and a shift in location of the maxima along $B$. The shift in $B$ is caused by $N_s$ increasing with $T$ as discussed in Supplementary Note 2. \textbf{b,} First maximum amplitude $R_{n=1}$ plotted as a function of $T$ for the spectra depicted in \textbf{(a)}. The black solid lines represent a fit to Eq.~(\ref{eq:decay_T}), demonstrating the dominant effect of MC scattering on TMF. \textbf{c,} Measured $\mathcal{l_\textrm{MC}}(T_c)$ plotted vs $T_c$ for Device1 (blue), Device2 (red) and Device3 (green) on double-logarithmic scale. The shaded regions depict the calculated values using Eq.~(\ref{eq:l_mc_T}) plotted vs $T$ with error bars. The grey region labeled GQ represents the theoretical value of $\mathcal{l_\textrm{MC}}$ from Eq.~(\ref{eq:GQ}) for comparison. A reference line $T^{-2}$ is drawn as a guide to the eye emphasizing that experimental $\mathcal{l_\textrm{MC}}(T_c)$ decay with $T^{-2}$. Experimental $\mathcal{l_\textrm{MR}}$ vs $T$ (orange) is plotted with a reference line depicting the expected $T^{-1}$ fall off. $\mathcal{l_\textrm{MR}} > \mathcal{l_\textrm{MC}}$ and $\mathcal{l_\textrm{MR}} > L_c$ throughout the experiments indicating the minimal effect of $\mathcal{l_\textrm{MR}}$ on TMF.}
\label{fig:fig3}
\end{figure}

The universal decay of $R_{n=1}(d_c, \mathcal{l_\textrm{MC}})$ provides an opportunity to measure $\mathcal{l_\textrm{MC}}$ experimentally. Figure~\ref{fig:fig3}a depicts the experimental TMF spectra measured at various $T$ for selected $L_c$ in Device1 ($L_c=$ 7 $\mu$m and 10.5 $\mu$m), Device2 ($L_c=$ 5 $\mu$m and 15 $\mu$m) and Device3 ($L_c=$ 2.6 $\mu$m and 12.8 $\mu$m). Measurements in other geometries can be found in Supplementary Note 5. We first convert the universal $R_{n=1}(d_c, \mathcal{l_\textrm{MC}})$ into $R_{n=1}(d_c, T)$, the quantity measured in experiments. Substituting $\mathcal{l_\textrm{MC}}(T)$ of Eq.~(\ref{eq:l_mc_T}) into Eq.~(\ref{eq:decay_l_mc}), we obtain: 
\begin{align}
R_{n=1}(d_c, T) = R_{n=1}(d_c, 0) - \Delta_{n=1}(d_c) \left(1 - \exp\left(-\left(\frac{T}{T_{c}}\right)^2\right)\right)
\label{eq:decay_T}
\end{align}
Equation~(\ref{eq:decay_T}) represents a model for $R_{n=1}(d_c, T)$ to which experimental data can be fit with three fitting parameters $-$ $R_{n=1}(d_c, 0)$, $\Delta_{n=1}(d_c)$ and most importantly, $T_{c}$. Knowing $T_{c}$ allows determination of $\mathcal{l_\textrm{MC}}(T = T_{c}) = (1.34 \pm 0.1) d_c$. By measuring $R_{n=1}(d_c, T)$ at various $L_c = d_c$ (each $L_c$ corresponding to one $T_{c}$), one can obtain $\mathcal{l_\textrm{MC}}$ at a series of temperatures $T = T_{c}$. We note that measuring $R_{n=1}(d_c, T)$ for different $T$ at fixed $L_c = d_c$ yields $\mathcal{l_\textrm{MC}}$ at a single $T = T_{c}$. Values for $\mathcal{l_\textrm{MC}}(T)$ for different $T$ can then be calculated by Eq.~(\ref{eq:l_mc_T}), but such values do not constitute a direct measurement. The scheme offers a straightforward interpretation in that measurements at the length scale $L_c$ probe e-e scattering at the energy scale set by $T_{c}$. Figure~\ref{fig:fig3}b depicts the experimental values for maxima $R_{n=1}(d_c, T)$ in Device1, Device2 and Device3 for $L_c = d_c$ of Fig.~\ref{fig:fig3}a. Notably, Fig.~\ref{fig:fig3}b shows that Eq.~(\ref{eq:decay_T}) is closely obeyed, providing evidence that Eq.~(\ref{eq:l_mc_T}) not only represents the most straightforward form compatible with phase-space arguments for a 2DES in a GaAs/AlGaAs heterostructure, but also is closely followed. Recent TMF experiments in graphene \cite{Lee2016, Berdyugin2020} have likewise obtained similar dependence on $T$ of maxima $R_n$. For every $L_c$, Fig.~\ref{fig:fig3}b yields corresponding $T_c$ and  $\mathcal{l_\textrm{MC}}(T_{c}) = (1.34 \pm 0.1) d_c$, plotted in Fig.~\ref{fig:fig3}c for all devices. Using Eq.~(\ref{eq:l_mc_T}), we can calculate the values of $\mathcal{l_\textrm{MC}}$ for continuous $T$. In Fig.~\ref{fig:fig3}c we overplot $\mathcal{l_\textrm{MC}}(T)$ calculated by Eq.~(\ref{eq:l_mc_T}) such that the shaded regions depict the possible range of $\mathcal{l_\textrm{MC}}$ for a particular device (see Supplementary Note 7 for sources of uncertainty in the calculation of $\mathcal{l_\textrm{MC}}$). Figure~\ref{fig:fig3}c shows that the data for $\mathcal{l_\textrm{MC}}(T_c)$ indeed decays as $T_{c}^{-2}$ (in accordance with Fermi liquid theory) which is an expected but nontrivial finding because rather than being an assumption it now stems from a direct measurement. Since the quantities $R_{n=1}(d_c, 0)$ and $\Delta_{n=1}(d_c)$ in Eq.~\ref{eq:decay_T} only depend on $B$ and not on $T$, they can be eliminated and a closed form expression can be obtained for $T_c$ as derived in Supplementary Note 6.

\section*{Discussion}

 The same slopes but different intercepts of $\mathcal{l_\textrm{MC}}$ vs $T$ in Fig.~\ref{fig:fig3}c show that in Device1, Device2 and Device3, $\mathcal{l_\textrm{MC}}$ follows a dependence $T^{-2}$, but with a multiplicative prefactor dependent on device. We surmise that the nonuniversal prefactor originates in device-dependent electrostatic environments for the 2DES. While all three devices are fabricated on nominally the same heterostructure, the 3D electrostatic environment of the 2DES can vary between devices due to residual charged impurities, leading to varying levels of dielectric screening \cite{kim2020}. We then expect $\mathcal{l_\textrm{MC}}$ to possess a nonuniversal prefactor dependent on the screening strength \cite{kim2020}, affecting the magnitude of $\mathcal{l_\textrm{MC}}$ but not the dependence $T^{-2}$. In Fig.~\ref{fig:fig3}c, the extracted $\mathcal{l_\textrm{MC}}$ is compared to values predicted by a commonly used theoretical expression \cite{GQ1982}: 
\begin{equation}
\mathcal{l_\textrm{MC}}^{-1} = \frac{(k_B T)^2}{hE_F v_F}\left\{\ln\frac{E_F}{k_BT} + \ln\frac{2q_\textrm{TF}}{k_F} +1\right\}
\label{eq:GQ}
\end{equation}
where $q_\textrm{TF}$ represents the Thomas-Fermi wavevector and $k_B$ the Boltzmann constant. The values of $T$, $E_F$, $v_F$, $k_F$, $q_\textrm{TF}$ are obtained from experiment. Agreement is observed between the data and calculated $\mathcal{l_\textrm{MC}}$ and Eq.~(\ref{eq:GQ}) in Device1 at lower $T$. However in Device2 and Device3 we observe longer $\mathcal{l_\textrm{MC}}$ than predicted by Eq.~(\ref{eq:GQ}), especially at lower $T$, and in all three devices the dependence on $T$ predicted by Eq.~(\ref{eq:GQ}) deviates from experiment. Again dielectric screening plays a role in the deviation. While many-body screening due to electrons is accounted for in Eq.~(\ref{eq:GQ}) using the random phase approximation (RPA), dielectric screening is not included, as noted in Ref.~\cite{kim2020}. Varying levels of dielectric screening and concomitantly varying 3D electrostatic environments hence explain the nonuniversal prefactor of $\mathcal{l_\textrm{MC}}$ as well as the deviation of Eq.~(\ref{eq:GQ}) from experiment. Notably, measurements of $\mathcal{l_\textrm{MC}}$ obtained from nonlocal resistance measurements at $B=$ 0 in Device3 \cite{Gupta2021}, yield similar values of $\mathcal{l_\textrm{MC}}$ as presented here, albeit with larger error bars.

\begin{figure}[!t]
\includegraphics[width=1\linewidth]{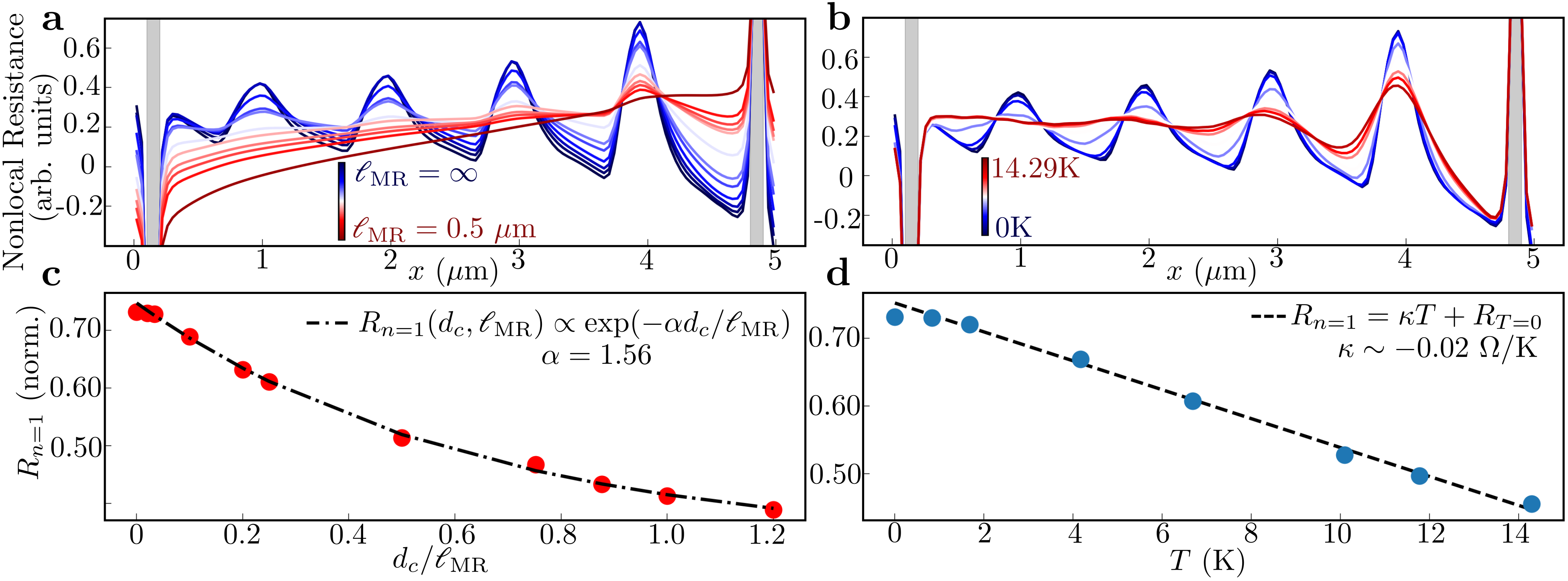}
\caption{\textbf{Decay of TMF with MR scattering and thermal Fermi surface broadening. a, b,} Simulated nonlocal resistance in T1 plotted versus position $x$ along the edge of the device into which the injection PC is placed, for variable $\mathcal{l_\textrm{MR}}$ from $\mathcal{l_\textrm{MR}} \rightarrow \infty$ to 0.5\ $\mu$m in \textbf{(a)}, and for variable $T$ from $T \rightarrow 0$ K to $T = 14.3$ K in \textbf{(b)}. \textbf{c,} Simulated $R_{n=1}(d_c, \mathcal{l_\textrm{MR}})$ assuming $\mathcal{l_\textrm{MC}} \rightarrow \infty$, plotted vs $d_c/\mathcal{l_\textrm{MR}}$ revealing an exponential decay with decay constant $\alpha=$ 1.56. \textbf{d,} Simulated $R_{n=1}$ assuming $\mathcal{l_\textrm{MR}}, \mathcal{l_\textrm{MC}} \rightarrow \infty$, plotted vs $T$ along with a linear fit (dashed line). Both MR scattering and thermal Fermi surface broadening play a lesser role in the decay of TMF in the experiments. }
\label{fig:fig4}
\end{figure}

We next address the effects on the decay of the TMF maxima of MR scattering due to impurity scattering and phonons, and of Fermi surface broadening under finite $T$ (Fig.~\ref{fig:fig4}) (broadening by injection energy has also been considered \cite{Hornsey1993, Williamson1990}). We first discuss MR scattering. Repeating the simulations, but now with only MR scattering present (variable $\mathcal{l_\textrm{MR}}$, $\mathcal{l_\textrm{MC}} \rightarrow \infty$), in Figs.~\ref{fig:fig4}a,c we find that $R_{n=1}(d_c, \mathcal{l_\textrm{MR}})$ follows the same exponential dependence as observed for MC scattering, albeit with a different $\alpha = 1.56$. While $\alpha$ is expected to be order $\mathcal{O}(\pi /2)$ as indeed borne out, the precise values for $\alpha$ follow simulation results. The same functional dependence is not surprising, since both MR and MC scattering result in the same physical process affecting TMF: randomization of individual carrier trajectories, leading to defocusing. The difference in $\alpha$ suggests that for same values of $\mathcal{l_\textrm{MC}}$ and $\mathcal{l_\textrm{MR}}$, the decay is slightly more sensitive to MC scattering than to MR scattering. However, throughout the range of $T$ in our experiment we have $\mathcal{l_\textrm{MR}} > \mathcal{l_\textrm{MC}}$ and $\mathcal{l_\textrm{MR}} > L_c$ due to the ultraclean heterostructure, and hence $\mathcal{l_\textrm{MR}}$ has minimal impact on the TMF spectra. This is corroborated in Fig.~\ref{fig:fig3}c, showing that the decay of $R_{n=1}(T)$ vs $T$ is not consistent with the measured $\mathcal{l_\textrm{MR}}(T)$ vs $T$ (obtained from standard 4-probe measurement of mobility and $N_s$). Figure~\ref{fig:fig3}c compares $\mathcal{l_\textrm{MC}}$ to $\mathcal{l_\textrm{MR}}$ and clearly shows $\mathcal{l_\textrm{MC}} < \mathcal{l_\textrm{MR}}$, ensuring the consistency of Eq.~\ref{eq:decay_l_mc} wherein MR scattering is ignored. We note that $\mathcal{l_\textrm{MR}}$ needs to be compared against $L_c$ and not $W$, while in contrast ballistic phenomena at $B=0$ require $\mathcal{l_\textrm{MR}} > W$ \cite{Gupta2021}. Therefore, the effect of MR scattering can be negated by choosing a sufficiently small $L_c$, which also allows for measurements of $\mathcal{l_\textrm{MC}}$ using higher $T_{c}$. 

To identify the effect of thermal broadening of the Fermi surface on the TMF signal, we perform ideal ballistic simulations ($\mathcal{l_\textrm{MR}}, \mathcal{l_\textrm{MC}} \rightarrow \infty$) at finite $T$. As indicated in Figs.~\ref{fig:fig4}b,d, thermal smearing merely produces a linear decay in the TMF amplitude, and is thus a subordinate effect compared to the effect of MR and particularly MC scattering. 

A surprising finding from the kinetic simulations lies in the presence of collective phenomena in a ballistic TMF setup - the formation of current vortices between the injector and collector probes (Fig.~\ref{fig:fig1}e), which causes a local enhancement in the magnetic field. Current vortices, usually associated with hydrodynamic intuition, have only recently been associated with the ballistic regime at $B=0$ \cite{chandra2019, chandraquantum, Gupta2021}, and the present work shows vortices can occur at finite $B$ as well. The vortices cannot be understood by examining individual single-particle trajectories, but rather appear as collective phenomena of all the particles as a whole, in currents resulting from a vector sum over all the trajectories. While TMF has seen extensive numerical study \cite{Hornsey1996, Ueta, Stegmann, Milovanovic, Beconcini, LeGasse}, current vortices have not been reported till this work. A probable reason lies in the observation that current computation may be dominated by shot noise in TMF simulations performed using particle schemes \cite{Milovanovic, Beconcini}, a limitation not suffered by the present high-resolution kinetic scheme. It would be interesting to check if vortices persist in the coherent transport regime \cite{LeGasse, Stegmann}.

To conclude, by combined experiments and high-resolution kinetic simulations we show that transverse magnetic focusing amplitudes decay exponentially due to electron-electron scattering, demonstrating the more general importance of electron-electron scattering in ballistic transport in high-mobility materials. Analysis of the transverse magnetic focusing amplitudes thereby allows for a precision measurement of the electron-electron scattering length, of importance in solid-state systems. The kinetic simulations reveal the hitherto unsuspected presence of current vortices even in a ballistic transverse magnetic focusing setup. 
 
\section*{Methods}
The geometries were patterned by gently wet etching of the GaAs/AlGaAs heterostructure in H$_2$SO$_4$/H$_2$O$_2$/H$_2$O solution after electron beam lithography, using PMMA as etching mask, to a depth removing the GaAs quantum well hosting the 2DES. Prior to electron-beam lithography, a Hall mesa was defined by photolithography and wet etching in the same solution. Ohmic contacts were annealed InSn. Measurements were performed at 4.2 K $< T<$ 36 K in a sample-in-exchange-gas system, using low frequency ($\sim$ 45 Hz) lock-in techniques under AC current bias without DC offsets. The transport properties of the unpatterned material were independently characterized on a sample in the van der Pauw geometry using the same methods but omitting lithography steps. 

\section*{Data Availability}
The data that support the plots within this paper and other findings of this study are available from the corresponding author upon reasonable request. 

\section*{Code Availability}
The source code for the simulations can be found at https://www.github.com/mchandra/bolt.

\section*{Acknowledgements}
A. G. and J. J. H. acknowledge support by the U.S. Department of Energy, Office of Basic Energy Sciences, Division of Materials Sciences and Engineering under Award No. DE-FG02-08ER46532 for the conceptualization of the experiments, device fabrication, measurements, data analysis and interpretation. The MBE growth and transport measurements at Purdue are supported by the U.S. Department of Energy, Office of Basic Energy Sciences, Division of Materials Sciences and Engineering under Award No. DE-SC0020138. S. F., G. C. G. and M. J. M. also acknowledge support from Microsoft Quantum. A. G., J. J. H., G. K. and M. C. acknowledge computational resources (GPU clusters Cascades and NewRiver) and technical support provided by Advanced Research Computing at Virginia Tech. J.J.H. acknowledges a publication subvention from VT OASF.

\section*{Author contributions}
A. G. and J. J. H. conceptualized and designed the experiments. A. G. performed the device fabrication and measurements. S. F., G. C. G. and M. J. M. provided the MBE grown high-mobility GaAs/AlGaAs heterostructure. G. K. and M. C. performed the kinetic simulations. A. G., J. J. H., G. K. and M. C. contributed to the data analysis and writing of the manuscript.

\section*{Competing interests}
The authors declare no competing interests.

\section*{Corresponding author}
Correspondence and requests for materials should be addressed to J. J. Heremans.

\ifarXiv


\section*{Supplementary material}

**Supplementary Note 1. Device fabrication and material properties**

The mesoscopic geometries were patterned using electron beam lithography followed by wet etching of the barriers, using PMMA as the etching mask. Each device contains several mesoscopic apertures (point contacts, PCs) separated by various distances $L_c$. Each PC can function either as current injector $i$ or collector (voltage detector) $c$. The PC resistance $R_{pc}$ varies between 450 Ω to 750 Ω at $T = 4.2$ K, depending on the PC and device. The devices were fabricated from GaAs/AlGaAs MBE-grown material hosting the two-dimensional electron system (2DES). The GaAs quantum well is located 190 nm below the surface, has a width of 26 nm, and is top- and bottom-doped by Si $\delta$−layers 80 nm removed from the quantum well and embedded in $Al_{0.32}Ga_{0.68}As$ barriers. Optimization of heterostructure design is described in Ref.[1].

The van der Pauw method was used to characterize electron transport properties of the unpatterned 2DES. The values of 2D resistivity $R_\square$ from the van der Pauw measurements and areal electron density $N_s$ from Hall measurements on the fabricated device, are used to obtain electron mobility $\mu$. At temperature $T = 4.2$ K, it is found that $N_s = 3 - 3.4 \times 10^{15}$ m$^{-2}$ (depending on device), and $R_\square = 2.75$ Ω/$\square$, yielding $\mu \approx 670 - 756$ m$^2$V$^{-1}$s$^{-1}$ (confirming the cleanliness of the material) and momentum relaxing mean-free path $\ell_{MR} = 65 - 69$ μm, calculated as $\ell_{MR} = v_F \tau_{MR}$. Here $v_F$ denotes the Fermi velocity, with $v_F = 2.21 - 2.72 \times 10^5$ m/s over the range $4.2\,\text{K} < T < 36\,\text{K}$, and $\tau_{MR}$ denotes the momentum relaxation time obtained from $\mu = e\tau_{MR}/m$ with $e$ the electron charge and $m = 0.067\,m_e$ the effective electron mass, with $m_e$ the free electron mass. The Fermi energy, $E_F = 10.1 - 11.2$ meV for $4.2\,\text{K} < T < 36\,\text{K}$. Non-parabolicity of the band structure was taken into account in calculating the transport properties[2,3]. $N_s$ (Supplementary Fig. 1a) and $R_\square$ (Supplementary Fig. 1b) increase with increasing $T$, while $\mu \sim 1/T$ (Supplementary Fig. 1c), as expected since $\mu$ is limited by scattering with acoustic phonons. Supplementary Figure 1d depicts $1/\mu$ vs $T$, indicating that $1/\mu(T) = 1/\mu_o + \alpha T$, where $\mu_o$ denotes $\mu$ limited by impurity scattering, $\alpha$ denotes a proportionality constant, and $\alpha T$ describes the linear dependence on $T$ due to (predominantly) acoustic phonon consistent with theory in the equipartition regime (4-40 K)[4]. The rate of MR scattering in the equipartition regime is approximated as[4]:

$$\frac{1}{\tau_{MR}} = \frac{1}{\tau_{ph}} + \frac{1}{\tau_{MR,o}} \approx A_{ph}T + \frac{1}{\tau_{MR,o}} \tag{S1}$$

where $1/\tau_{ph}$ represents the phonon scattering rate, $A_{ph}$ the phonon scattering coefficient, and $1/\tau_{MR,o}$ the residual scattering rate due to impurities[5]. By plotting $1/\tau_{MR}$ (with $\tau_{MR}$ calculated from experimental values of $N_s(T)$ and $\mu(T)$ (Supplementary Fig. 1d), we indeed find that Eq. (S1)

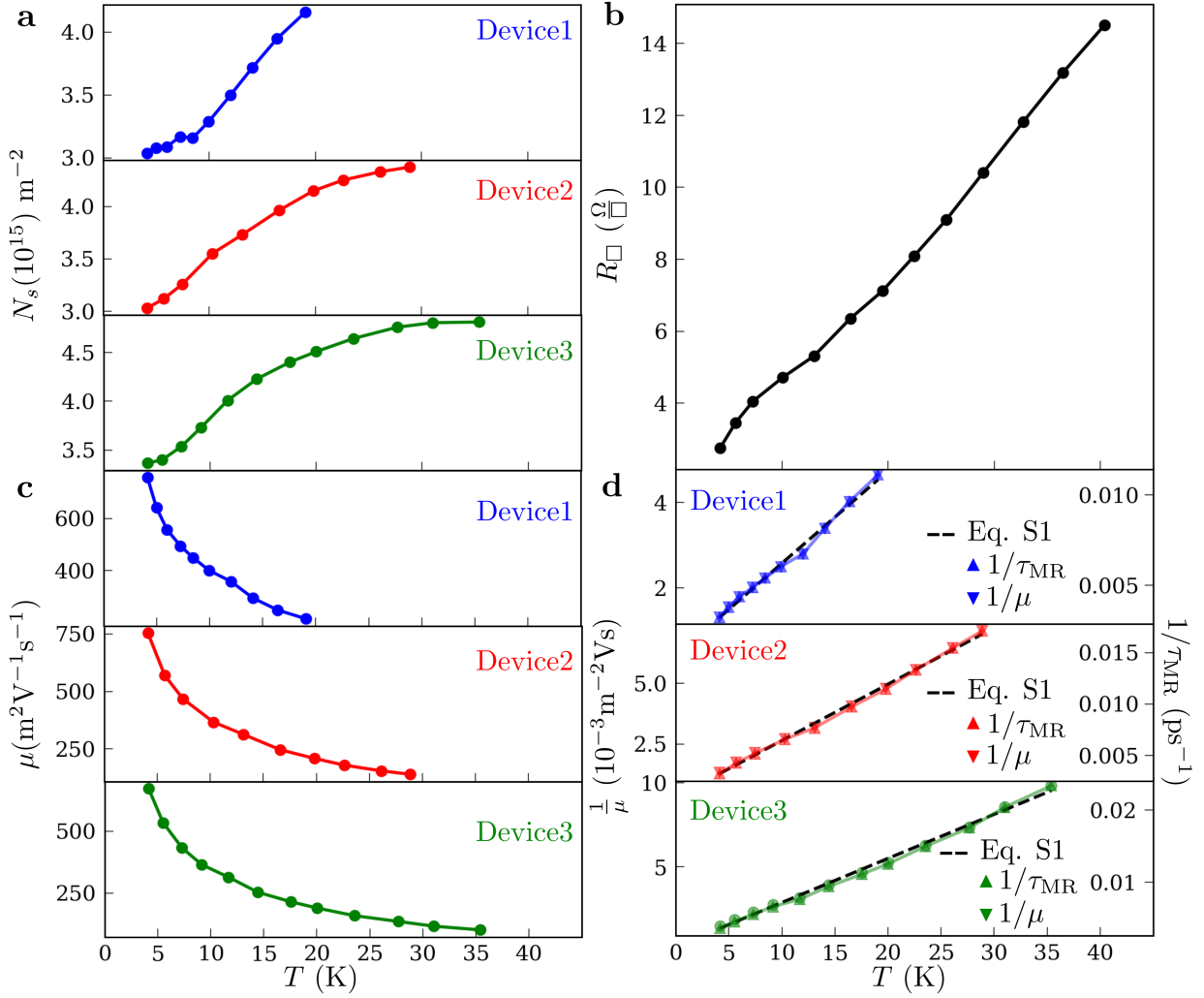

**Supplementary Figure 1 | Transport characteristics. a,** Carrier density $N_s$ vs $T$ for Device 1 (blue), Device2 (red) and Device3 (green). **b,** 2D resistivity $R_\square$ vs $T$ from van der Pauw measurements on the unpatterned material. **c,** Mobility $\mu$ vs $T$. **d,** $1/\mu$ vs $T$ (left axis) and $1/\tau_{\mathrm{MR}}$ vs $T$ (right axis). From the linear fit (black dashed line) and Eq. (S1), we extract $A_{ph} \approx 5.1 - 6.2 \times 10^8$ s$^{-1}$K$^{-1}$ and $\tau_{\mathrm{MR,o}} \approx 9.1 - 13.8 \times 10^{-10}$ s depending on the device.

describes the dependence on $T$ of $1/\tau_{\mathrm{MR}}$ well and that acoustic phonon scattering dominates the MR scattering in the range of $T$ of the experiments, as expected for a high-$\mu$ 2DES.

**Supplementary Note 2.   Properties of TMF**

Due to expected reciprocity relations[6,7], TMF results should be symmetric on exchanging the injector $i$ and collector $c$ and changing the polarity of magnetic field $B$ applied normal to the 2DES plane. Supplementary Figure 2a shows an example of this relation for Device1. While discrepancies

3

can arise in mesoscopic devices[8], in the present devices these are small, allowing us to show the TMF spectra for only one polarity of $B$. The cyclotron diameter $d_c$ approaches the conducting aperture width $w \approx 0.6\,\mu$m at $B = 0.3$ T, and hence well-defined semiclassical cyclotron orbits reflecting off the barrier require $B < 0.3$ T. As Supplementary Fig. 2b illustrates, the relevant data in Device1 hence occurs for $B \leq 0.2$ T (for other devices, the range of $B$ can be different depending on $L_c$). Supplementary Figure 2b also illustrates the magnetoresistance background often superposed on TMF (dashed red line). The magnetoresistance background occurs independently of the TMF and can have several origins. The background can be carefully identified using smoothing filters (Ref.[9] uses a Gaussian smoothing filter) and then be removed. In this work, however, for one-to-one correspondence with the simulations, we do not remove the background and present the untreated

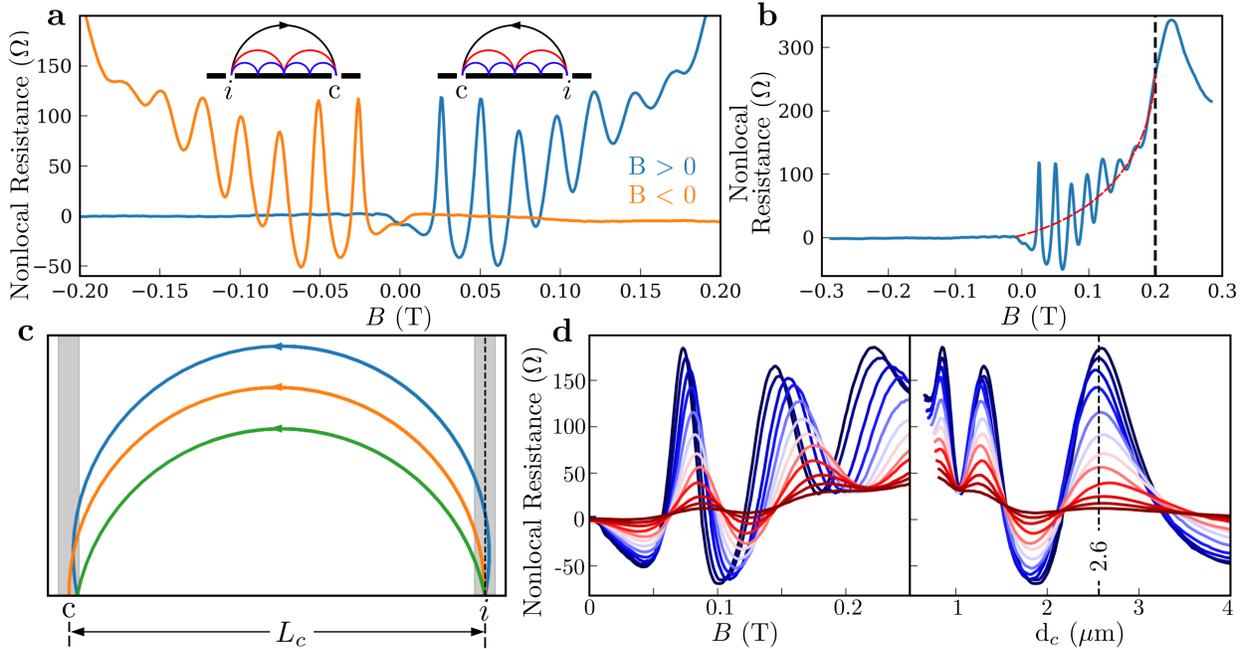

**Supplementary Figure 2 | Properties of TMF. a,** TMF at $T = 4.2$ K for D1 ($L_c = 7$ $\mu$m) when current and voltage contacts are exchanged and polarity of $B$ inverted, illustrating reciprocity in resistance. The insets depict the semiclassical orbits ($\pi/2$ injection) corresponding to the TMF maxima. **b,** $B > 0$ spectra in (a) illustrating that the relevant data lies at $B < 0.2$ T. The dashed red line depicts the positive background magnetoresistance superposed on TMF. **c,** TMF geometry and cyclotron orbits, illustrated with the injection angle differing from $\pi/2$ to the barrier (orange orbit's injection angle at $\pi/2$ as reference). The grey vertical bars depict the injector and collector PC apertures. **d,** left panel: TMF spectra for Device3 ($L_c = 2.6$ $\mu$m) illustrating a shift vs $B$ as $T$ is increased. Right panel: the shift vanishes and the TMF spectra coincide when plotted vs $d_c$.

.

4

experimental signal.

Supplementary Figure 2c illustrates that a cyclotron orbit starting at an angle different from $\pi/2$ to the barrier, will land in the vicinity of $c$ at a distance less than $L = d_c$ from $i$ (illustrated case corresponds to first maximum in the TMF spectrum). However, if the injection angles are not far from $\pi/2$, the orbits undergo magnetic focusing onto $c$, a property due to $d_c$ being an extremal length scale of the cyclotron orbit for a GaAs 2DES. More generally, in a semiclassical approach, under $B$ the path in reciprocal space coincides with cross-sections of the Fermi surface corresponding to equal-energy contours[10,11]. In real space the cyclotron orbit corresponds to a path of the same shape, rotated by $\pi/2$ and scaled as $1/B$. TMF spectra emphasize those orbits corresponding to extremal Fermi surface cross-sections. If the Fermi surface is circular with diameter $2k_F$ (as for a GaAs 2DES), then $d_c = 2\hbar k_F/eB$ (where $k_F$ represents the Fermi wave vector). Carrier reflection from a potential barrier obtained by gentle wet etching is predominantly specular and hence TMF maxima appear when $L_c = nd_{cn} = 2\hbar k_F/eB_n$ where $n$ is an integer and $B_n$ is the magnetic field corresponding to the n$^{th}$ maximum (inset in Supplementary Fig. 2a). However, as discussed in the main text, due to spread in the injection angles, maxima will appear at $B$ slightly lower than $B_n$. In Supplementary Figs. 2a-b, the TMF amplitude decreases with increasing $n > 3$ and for $n > 6$, the maxima lose definition. Two reasons contribute to this effect: 1) at higher $B$, $d_c \lesssim w$ and the semiclassical cyclotron orbits then do not have a clear point or angle of origin in the injector, and 2) as discussed in the main text, the spread in injection angles leads to a gradual defocusing after several skipping events off the barrier, even under perfect specular reflection off the barrier.

As illustrated in Fig. 3a main text, a shift in $B$ in the TMF spectra and hence at the location of TMF maxima is observed as $T$ is increased. The shift is due to increasing $N_s$ with $T$ (Supplementary Fig. 1a) leading to an increase in $k_F = \sqrt{2\pi N_s}$ with $T$. Since the position of $c$ in the experiments remains fixed ($L_c = nd_{cn}$ remains fixed), $B_n = 2\hbar k_F/(e\, nd_{cn})$ increases proportionately with increasing $k_F$. Using $L_c = 2.6$ $\mu$m in Device3, Supplementary Fig. 2d illustrates that when the TMF specta are plotted vs $B$ the shift is clearly visible, while the shift vanishes when the TMF spectra are plotted vs $d_c$.

**Supplementary Note 3.** **Effects of $\ell_{MC}$ and $d_c$ on the current streamlines and voltage profiles**

Here we depict simulation results of the spatial current streamlines and voltage profiles on the theoretical test device T1 (5 $\mu$m × 5 $\mu$m square device), with current injection from the right edge and extraction from the left edge (Supplementary Fig. 3). We also depict additional

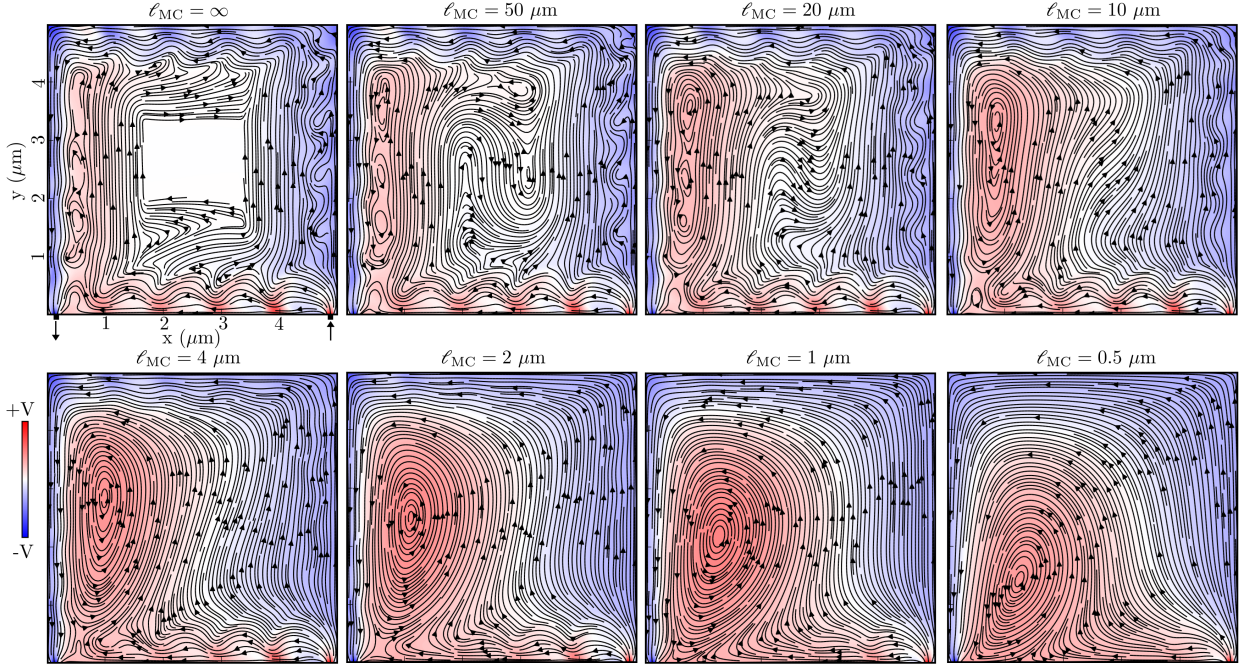

**Supplementary Figure 3 | Effect of $\ell_{\mathrm{MC}}$ on TMF.** Current streamlines and voltage contour plots for T1 at several values of $\ell_{\mathrm{MC}}$ with fixed $d_c = 1$ $\mu$m. Starting with $\ell_{\mathrm{MC}} \to \infty$, we observe a ballistic flow profile with skipping cyclotron orbits propagating along the edges of the device. As $\ell_{\mathrm{MC}}$ approaches the device scale, a vortex forms as expected when the system approaches the hydrodynamic regime (short $\ell_{\mathrm{MC}}$). When $\ell_{\mathrm{MC}} \lesssim d_c = 1\,\mu$m, the cyclotron orbits are suppressed by the device scale vortex. With $\ell_{\mathrm{MC}}$ the shortest length scale, the system is then in the hydrodynamic regime.

simulation results of the current streamlines and voltage profiles on Device1 (Supplementary Fig. 4). The test devices T1 and T2 (10 $\mu$m × 7.5 $\mu$m rectangular device) help develop an intuition for results in more complex experimental geometries, such as Device1. In Fig. 2b main text we depicted universal curves for a normalized $R_{n=1}$ vs parameter $d_c/\ell_{\mathrm{MC}}$ for geometry T1 (for $L_c = d_c = 0.5, 1.0\,\mu$m; variable $\ell_{\mathrm{MC}}$), for geometry T2 (for $L_c = d_c = 1.5, 2.5\,\mu$m; variable $\ell_{\mathrm{MC}}$), and for Device1 ($L_c = d_c = 7\,\mu$m; variable $\ell_{\mathrm{MC}}$). We showed that the normalized curves for all the devices overlap, such that $R_{n=1}$ follows an exponential decay vs $d_c/\ell_{\mathrm{MC}}$ independent of geometry.

In the simulations we set $\ell_{\mathrm{MR}} \to \infty$. In Supplementary Fig. 3 we visualize the effect of the variable $\ell_{\mathrm{MC}}$ on the current streamlines and voltage profiles in T1 at a fixed $d_c = 1\,\mu$m (corresponding to the blue curve in Fig. 2b main text for $L_c = d_c = 1.0\,\mu$m). Salient observations are a ballistic flow profile with skipping cyclotron orbits along the edges of the device persisting for $\ell_{\mathrm{MC}} \gtrsim 4\,\mu$m, and a vortex forming and consuming the cyclotron orbits as $\ell_{\mathrm{MC}}$ approaches first the device scale and then $d_c$ (in the hydrodynamic regime). In Supplementary Fig. 4, we visualize the effect of

6

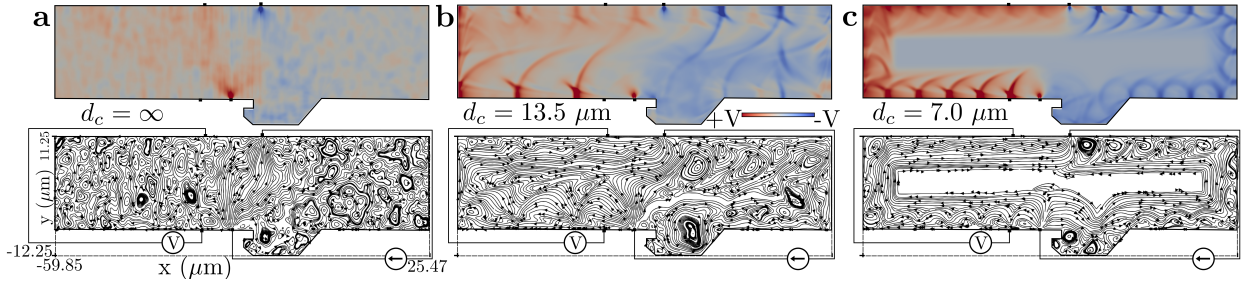

**Supplementary Figure 4 | Effect of $d_c$ (or $B$) on TMF a,** Current streamline and voltage contour plots for Device1 at $B = 0$, where the device exhibits numerous current vortices forming at various device scales due to (specular) scattering at the device boundaries[12–14]. **b,** Current streamline and voltage contour plots for small $B = 0.013$ T, corresponding to $d_c = 13.5$ $\mu$m. The current lines are more streamlined. Yet, since $d_c = 13.5$ $\mu$m approaches the device scale $W \sim 21$ $\mu$m, scattering at the device boundaries still influences the current streamlines and voltage contours, as discussed in Supplementary Note 4. **c,** Current streamline and voltage contour plots for $B = 0.025$ T, corresponding to $d_c = 7.0$ $\mu$m. The local net current flows follow the device boundaries, minimizing scattering at boundaries.

variable $d_c$ (hence variable $B$) on the current streamlines and voltage profiles in Device1 in the ballistic limit ($\ell_{\mathrm{MC}} \to \infty$). A higher $B$ (lower $d_c$) encourages larger-scale features in the current streamlines, suppressing the wide range of sizes observed for current vortices at $B = 0$, ultimately tending to the local net current flows following the device boundaries.

**Supplementary Note 4. Effect of $d_c \sim W$**

A precondition for an accurate measurement of $\ell_{\mathrm{MC}}$ from the TMF signal is $d_c < W$, where $W$ represents the device scale. When $d_c \sim W$, (specular) boundary scattering affects the TMF signal (as observed in Supplementary Fig. 4b). The decay of $R_{n=1}$ with $\ell_{\mathrm{MC}}$ then deviates from the universal curve, as mentioned in the main text. We exemplify this deviation in Supplementary Fig. 5 using $d_c = 2$ $\mu$m in the T1 device with $W = 5$ $\mu$m ($\ell_{\mathrm{MR}} \to \infty$ in the simulations). We note that the deviation is particularly noticeable near the ballistic limit ($\ell_{\mathrm{MC}} \to \infty$), where the dominant scattering occurs with device boundaries. In contrast, with strong MC scattering (short $\ell_{\mathrm{MC}} < W$), the deviation is subdued owing to $\ell_{\mathrm{MC}}$ and not $W$ being the shortest length scale in the system. We point out that despite this expected deviation from universality, one can still obtain values for $\alpha$ by justifiably not considering the deviant data points in the curve fitting.

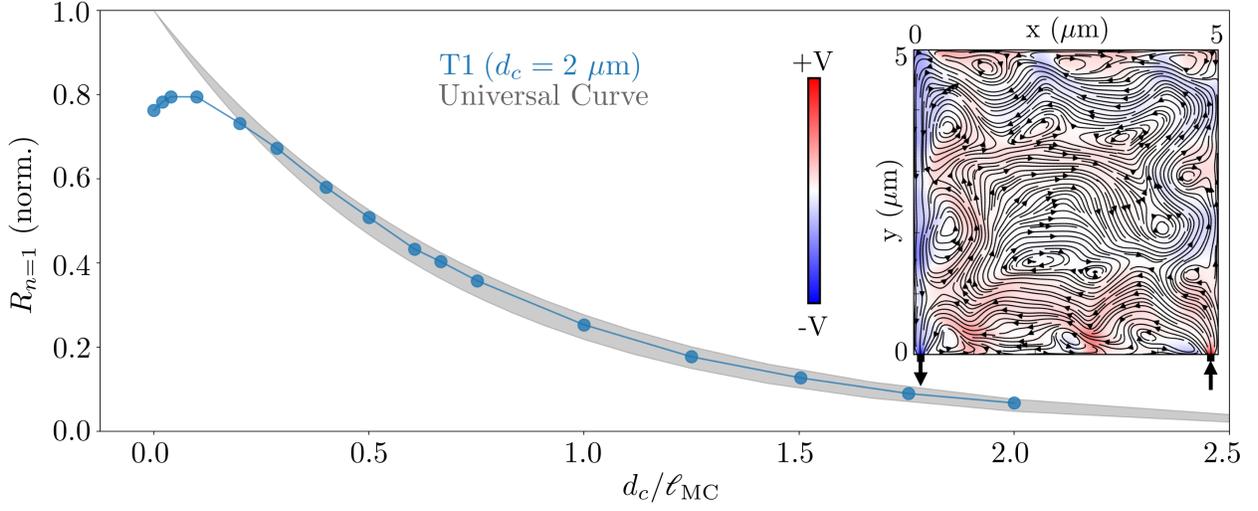

**Supplementary Figure 5 | Effect of $d_c \sim W$.** $R_{n=1}$ vs $d_c/\ell_{\rm MC}$ for $d_c = 2$ $\mu$m in geometry T1. The deviation from the universal curve (shaded grey region) at long $\ell_{\rm MC}$ (close to the ballistic limit) can be attributed to boundary scattering effects originating from the condition $d_c \ll W$ being violated. The inset depicts the current streamlines and voltage contours at $d_c = 2$ $\mu$m for $\ell_{\rm MC} \to \infty$ (ballistic limit). Comparing to the first panel of Supplementary Fig. 3 (with $d_c = 1$ $\mu$m and $\ell_{\rm MC} \to \infty$) where the current flow is restricted to boundaries with little effect from parallel or perpendicular boundaries, a clear difference is visible with the boundaries affecting the current streamlines.

**Supplementary Note 5.   Experimental TMF spectra for remaining $L_c$**

The hitherto undepicted experimental TMF spectra for distances $L_c$ in Device2 and Device3, used for estimating $\ell_{\rm MC}(T_c)$ (Fig. 3c, main text), are depicted in Supplementary Fig. 6. In Device2, $L_c$ ranges from 3 $\mu$m to 15 $\mu$m with $W$= 42 $\mu$m. In Device3, $L_c$ ranges from 1.3 $\mu$m to 20.5 $\mu$m with $W$= 24 $\mu$m. We omitted the longer $L_c$ >12.8 $\mu$m in Device3 because of boundary effects as discussed in the previous section. Supplementary Note 2 (Supplementary Fig. 2d) discusses the observed upward shift in $B$ for the location of maxima as $T$ increases, due to increasing $N_s$.

**Supplementary Note 6.   A closed form expression for $T_c$**

By eliminating the $B$ dependent quantities $R_{n=1}(d_c, 0)$ and $\Delta_{n=1}(d_c)$ in Eq. 4 main text, we derive a closed form expression for $T_c$. Rewriting Eq. 4 with $R_{n=1}$ measured at $T = T + \delta T$, we obtain:

8

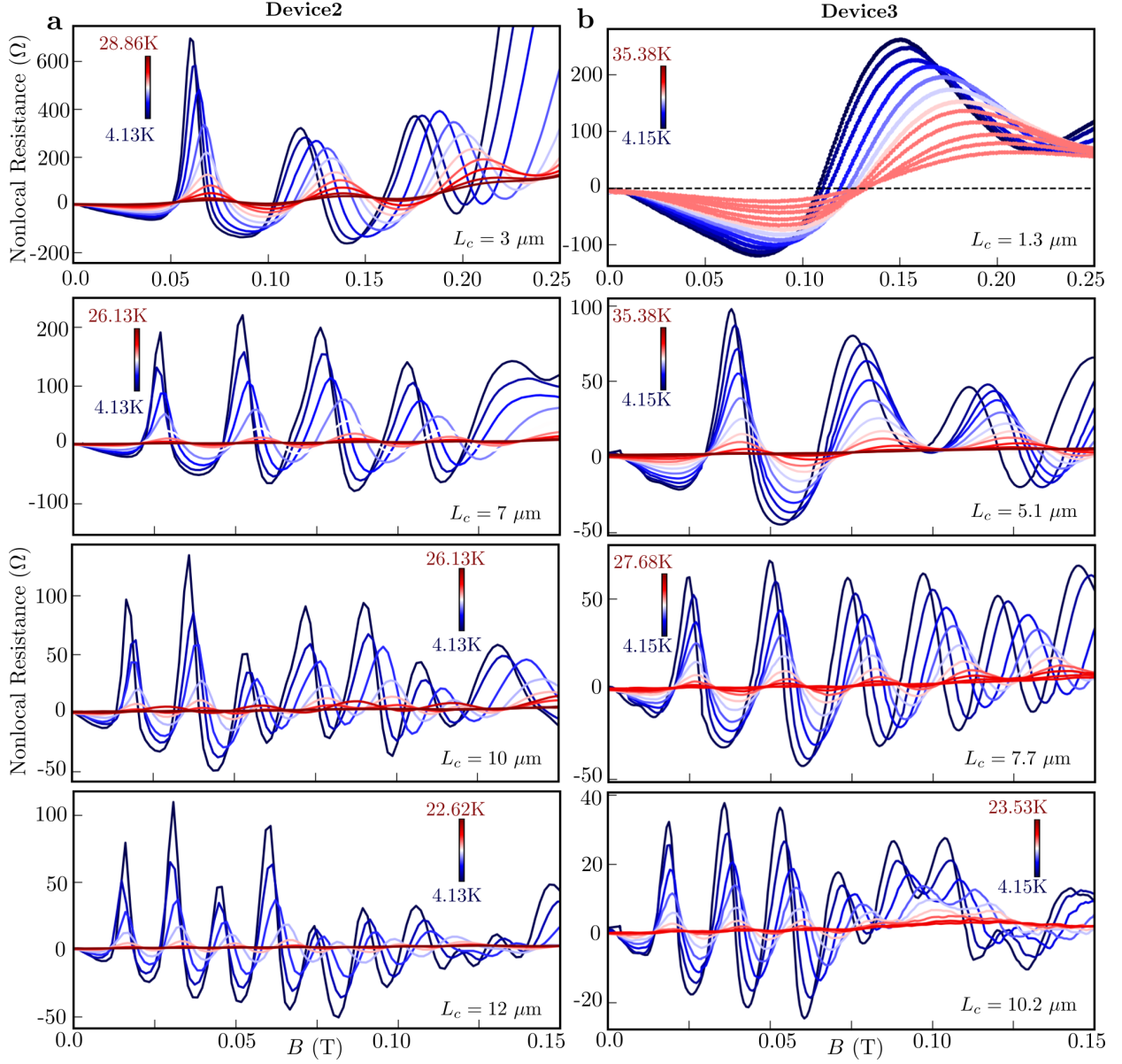

**Supplementary Figure 6 | Experimental TMF spectra a,** Experimentally measured TMF spectra for $L_c = 3, 7, 10, 12$ $\mu$m in Device2. **b,** Experimentally measured TMF spectra for $L_c = 1.3, 5.1, 7.7, 10.2$ $\mu$m in Device3. The range of $T$ for each $L_c$ is indicated. We observe an upward shift in $B$ for the location of maxima with increasing $T$, due to increasing $N_s$ (Supplementary Fig. 2d).

$$R_{n=1}(d_c, T + \delta T) = R_{n=1}(d_c, 0) - \Delta_{n=1}(d_c)\left(1 - \exp\left(-\left(\frac{T+\delta T}{T_c}\right)^2\right)\right) \tag{S2}$$

Subtracting Eq. 4 main text from the above, we eliminate one of the non-universal parameters, $R_{n=1}(d_c, 0)$:

$$\delta R_{n=1}(d_c, T) = R_{n=1}(d_c, T + \delta T) - R_{n=1}(d_c, T)$$
$$= \Delta_{n=1}(d_c) \left( \exp\left(-\left(\frac{T + \delta T}{T_c}\right)^2\right) - \exp\left(-\left(\frac{T}{T_c}\right)^2\right) \right) \quad (S3)$$

Using the first order Taylor expansion, we write $\delta R_{n=1}(d_c, T)$ as:

$$\delta R_{n=1}(d_c, T) = -\Delta_{n=1}(d_c) \frac{2T \delta T}{T_c^2} \left( \exp\left(-\left(\frac{T}{T_c}\right)^2\right) \right) \quad (S4)$$

Dividing $\delta R_{n=1}(d_c, T_1)$ and $\delta R_{n=1}(d_c, T_2)$ measured at two distinct temperatures $T = T_1$ and $T = T_2$, we eliminate the other non-universal parameter $\Delta_{n=1}(d_c)$:

$$\frac{\delta R_{n=1}(d_c, T_1)}{\delta R_{n=1}(d_c, T_2)} = \frac{T_1 \delta T_1}{T_2 \delta T_2} \left( \exp\left(\frac{T_2^2 - T_1^2}{T_c^2}\right) \right) \quad (S5)$$

Rearranging and, taking the natural log on both sides,

$$ln\left( \frac{\delta R_{n=1}(d_c, T_1)}{T_1 \delta T_1} \frac{T_2 \delta T_2}{\delta R_{n=1}(d_c, T_2)} \right) = \left( \frac{T_2^2 - T_1^2}{T_c^2} \right) \quad (S6)$$

In the limit $\delta T \to 0$, we obtain the closed form expression for $T_c$:

$$T_c^2 = -\frac{T_2^2 - T_1^2}{\ln\left(\frac{1}{T} \mid \frac{\partial R_{n=1}}{\partial T} \mid\right)\big|_{T_2} - \ln\left(\frac{1}{T} \mid \frac{\partial R_{n=1}}{\partial T} \mid\right)\big|_{T_1}} \quad (S7)$$

where $\partial R_{n=1}/\partial T$ is the temperature slope of $R_{n=1}$. Equation S7 shows that in principle $T_c$ can be obtained without fitting procedure. We note that for the Eq. S7 to be valid, the experimental data must obey Eq. 4 main text, a necessary check that must be performed.

**Supplementary Note 7.** Fitting procedures and sources of uncertainty in the estimation of $\ell_{\mathbf{mc}}$

In Eq. 4 main text, $R_{n=1}(d_c, T = 0)$, $\Delta_{n=1}(d_c)$ and $T_c$ are the three fitting parameters to the experimental $R_{n=1}(d_c, T)$ vs $T$ data.

Uncertainties in the calculation of $\ell_{\mathrm{MC}}$ using the method described in this work result from the following sources :

1. Uncertainty in the determination of the universal decay constant $\alpha$, which could result from the following:

    (a) Curve fitting error: Error in the determination of $\alpha$ by fitting the simulated decay curve ($R_{n=1}(d_c, \ell_{\text{MC}})$ vs $\ell_{\text{MC}}$) for a fixed geometry at fixed $B$ while varying $\ell_{\text{MC}}$, to Eq. 1, main text. This error can be reduced by using a large number of closely-spaced simulated data points in the decay curve, and ensuring that each simulation has run for a long enough time to reach steady state solution, both resulting in a better fit. The maximum fitting error is found to be 0.039 (obtained from simulation data in different device geometries under several representative $B$). We note that this uncertainty is smaller than the statistical uncertainty described below.

    (b) Statistical uncertainty: Spread in the estimation of $\alpha$ obtained by fitting Eq. 1 main text to $R_{n=1}(d_c, \ell_{\text{MC}})$ vs $\ell_{\text{MC}}$ curves obtained from simulations, each with different $B$ and/or device geometries. As mentioned in the main text, the value and associated statistical uncertainty in the simulations are found as $\alpha = 1.34 \pm 0.1$.

    (c) Position of voltage counterprobe: We note that placing the voltage counterprobe ($V-$) too near the injector PC (where cyclotron orbits may impinge on it) can introduce a systematic error in the estimation of $\alpha$. The universal value $\alpha = 1.34 \pm 0.1$ is found when the counterprobe is placed at a faraway location where the reference voltage remains constant as a function of $T$ and $B$. In the simulated test geometries T1 and T2, we achieve this by keeping the counterprobe at zero potential, which is defined to be the potential at which the carrier density is the same as the unperturbed equilibrium carrier density (see for example, white region in center of first panel in Supplementary Fig. 3). In Device1, for one-to-one correspondence with the experiments, we place the counterprobe in the simulations at the same location as in the experiments, which yields $\alpha = 1.26$ (inset of Fig. 2a, main text). However, as illustrated in Supplementary Fig. 7, in Device1 we find that $\alpha$ varies as the position of the counterprobe is moved horizontally ($\|x$ in Supplementary Fig. 7) along the top edge of the device. The variation decreases as we move the counterprobe to the right owing to the carriers from the injector PC having to traverse a larger distance before impinging on the counterprobe (Supplementary Fig. 7a). We conclude that placing the counterprobe faraway from the injector PC (at a distance $\gg d_c$) leads to a lower error in the experimental estimate of $\alpha$ and hence of $\ell_{\text{MC}}$. Another approach for minimizing the variation in $\alpha$ consists of increasing the conducting width of the counterprobe apertures, which averages out po-

tential variations within the counterprobe width (Supplementary Fig. 7b-c). We make use of both approaches in measurements on Device2 and Device3, where counterprobes of large width $\sim 20\,\mu$m are separated by up to mm (hence many $d_c$ and many $\ell_{\mathrm{MR}}$) from the injector PCs.

2. Experimental curve fitting uncertainty: The maximal % uncertainty, across three devices, in the fitting parameters $R_{n=1}(d_c, T=0)$ and $\Delta_{n=1}(d_c)$ obtained by fitting Eq. 4, main text to the experimental $R_{n=1}(d_c, T)$ vs $T$ data, are 8.5% and 8.7% respectively. The uncertainty in fitting parameter $T_c$ is less than 5.9%.

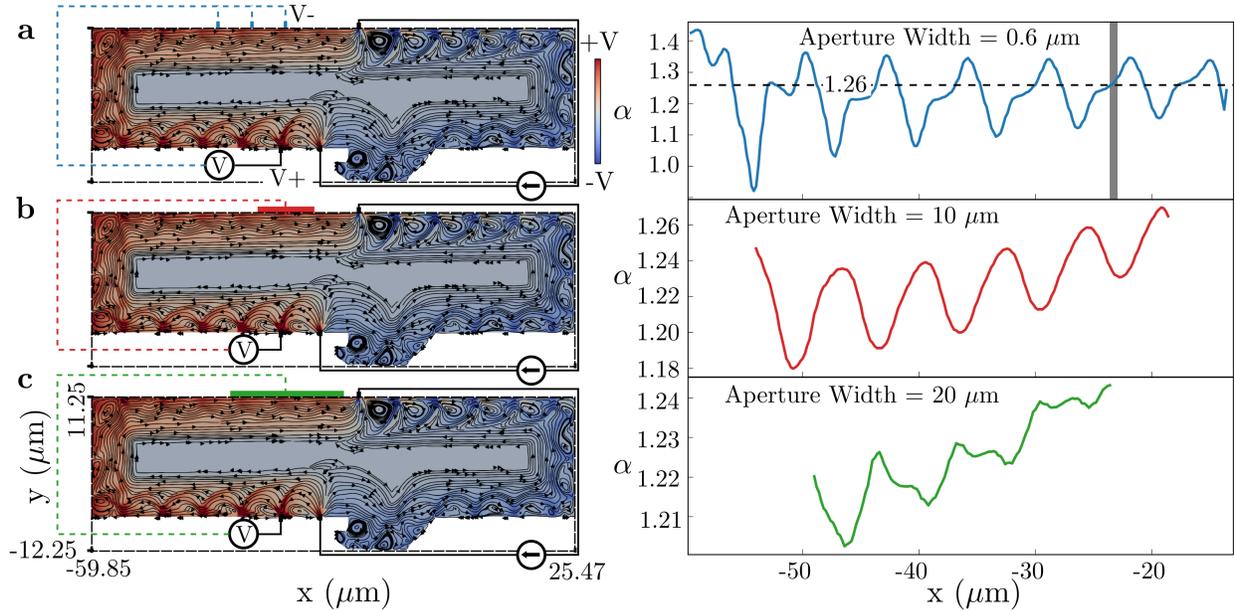

**Supplementary Figure 7 | Variation of $\alpha$ with voltage counterprobe position. a,b,c.** Current streamlines and voltage contour plots for Device1, when the position of the voltage counterprobe ($V-$), of aperture width **(a)** $0.6\,\mu$m, **(b)** $10\,\mu$m and **(c)** $20\,\mu$m, is varied along $x$ on the top edge of the device (shown schematically with dotted blue lines in **(a)**). The right panels depict $\alpha$ vs the position $x$ of the counterprobe for the corresponding counterprobe width in the left panel. The variation in $\alpha$ decreases on increasing the distance that skipping orbits originating from the injector have to travel along the boundary (counterclockwise in this case) to reach the counterprobe, and on increasing the width of the counterprobe aperture. The actual position of the counterprobe in the experiments is indicated by the grey bar in the right panel of **(a)**.


\fi

\end{document}